\documentclass[twocolumn, amsmath, stfloats, superscriptaddress, amsfonts,prb]{revtex4}
\pdfoutput=1
\usepackage{graphicx}
\usepackage{subfigure}
\usepackage{mathtools}

\usepackage{threeparttable}
\usepackage{longtable}
\usepackage{float}
\usepackage{color,soul}
\usepackage{hyperref}
\usepackage{breakurl}
\makeatletter
\def\url@leostyle{%
  \@ifundefined{selectfont}{\def\UrlFont{\sf}}{\def\UrlFont{\small\ttfamily}}}
\makeatother
\begin{document}
\title{The statistics of tin whisker diameters versus the underlying film grains}
\author{O. A. Oudat}\email{osama.oudat@rockets.utoledo.edu}\affiliation{Department of Physics and Astronomy, University of Toledo, Toledo,OH 43606, USA}
\author{Vamsi Borra}\email{vamsi.borra@utoledo.edu}\affiliation{Department of Electrical Engineering and Computer Science, University of Toledo, Toledo, OH 43606, USA}
\author{Daniel G. Georgiev}\email{daniel.georgiev@utoledo.edu}\affiliation{Department of Electrical Engineering and Computer Science, University of Toledo, Toledo, OH 43606, USA}
\author{V. G. Karpov}\email{victor.karpov@utoledo.edu}\affiliation{Department of Physics and Astronomy, University of Toledo, Toledo,OH 43606, USA}

\begin{abstract}
We compare the statistics of tin whisker diameters to that of the underlying film grains. Both are well approximated by the lognormal distributions. However, the parameters of those distributions can be rather different, not confirming the assumption that each whisker grows from a single grain. We conclude that several adjacent grains with similar crystal orientations can contribute to a whisker development. Our observations are consistent with the recent theory of multi-filament whisker structure. A  modification of the particle size log-normal distribution is developed clarifying the nature of its dispersion.

\end{abstract}

\maketitle

\section{Introduction}\label{sec:intro}
Hair like protrusions called metal whiskers (MWs) spontaneously formed on surfaces of many metals, present a significant threat to many technologies. \cite{NASA1,barnes,galyon2003,brusse2002} Their underlying physics remains poorly understood. \cite{panashchenko2009,panashchenko2012,borra2018,karpov2014,karpov2015} MWs exhibit significant statistical variations of their lengths \cite{fang2006,panashchenko2009,panashchenko2012,susan2013} and characteristic diameters, \cite{panashchenko2009,meschter2015} mutually uncorrelated, both fit well by the log-normal distributions. While the nature of lengths variations was addressed, \cite{karpov2014,niraula2015,subedi2017} the origin of variations in whisker diameters remains uncharted territory.

MW concentrations are small compared to the surface grain concentration (by a factor of $10^{-3}-10^{-5}$) varying exponentially between different local regions on the surface; \cite{galyon2003,brusse2002,davy2014,zhang2004,tu2005,bunian2013} some of the nominally identical samples may exhibit no MW, others showing significant MW infestations. Multiple published observations show that MW shapes exhibit rather irregular cross-sections including even hollow MWs. It was argued \cite{borra2018} that the MW cross-section shape is determined by that of the underlying charge patches giving rise to MW through the electrostatic mechanism.

A broadly shared hypothesis is that MW grow from certain rare grains possessing uncommon structure parameters; attempts were made to identify such grains. \cite{jagtap2017,pei2012,kakeshita1982} At least partially, that hypothesis is based on the observations that MWs do not significantly change their diameters since conception, \cite{panashchenko2009} which could be attributed to the grain boundary confinement. Another argument in favor of that hypothesis is the log-normal grain diameter statistics found for a variety of different materials. \cite{granquist1976,pande1987,vaz1988,kiss1999}
It was hypothesized \cite{kakeshita1982} that MW diameters significantly exceeding that of grains can be explained by the grain recrystallization while forming MW that increases the `founding' grain to the actual MW diameter.

Here, we experimentally determine and compare the statistics of MW and grain characteristic diameters for two different types of Sn films (Section \ref{sec:exp}). We then discuss a possible nature of the observed log-normal distributions and related implications for the physics of MW formation (Section \ref{sec:disc}). Also, we revisit the applicability of log-normal distribution to particle size description and propose a modification that clarifies the physical meaning of its dispersion that is shown to be size dependent.  Our conclusions are presented in Section \ref{sec:conl}.

\section{Experimental Details}\label{sec:exp}
\subsection{Sample preparation}\label{sec:samples}
We used Sn film samples deposited by two techniques.\\
(1) The evaporated samples were made using a Denton vacuum thermal evaporator (DV-502A turbo auto high vacuum evaporator) and Sn pellets (from Kurt J. Lesker) of 99.999\% purity. Following the recipe described in earlier publications, \cite{vasko2015a,killefer2017,borra2016} we used 3 mm thick Pilkington TEC-15 glass (soda lime glass, coated with Fluorine doped Tin Oxide with a sheet resistance of 15 $\Omega/\square$) as a substrate. The film thickness was close to 250 nm as determined with a  quartz crystal microbalance thickness monitor.

(2)The electroplated samples were deposited on mechanically polished copper coupons. After washing the coupons to remove organic traces, they were placed in an electroscrub bath. The bath solution was maintained at ~60 °C.  Using a galvanostat, a current of 200 mA was applied for 30 s, then reversed for 10 s, and reversed again for 30 s. The sample was then rinsed with distilled water and placed into a metal activator bath for 2 min to remove the oxide layer and ensure a clean metal-to-metal bond with the Sn film. Finally, the coupons were submerged into a sulfuric acid based electroplating solution where  a current of 200 mA was passed through the solution for a time commensurate with the desired thickness of the tin film. A large 99.95\% pure tin foil, which was submerged into the bath and connected to the anode of the galvanostat, was used as an electrode.


\subsection{Imaging and diameter measurements}\label{sec:image}
\begin{figure}[h!]
\includegraphics[width=0.37\textwidth]{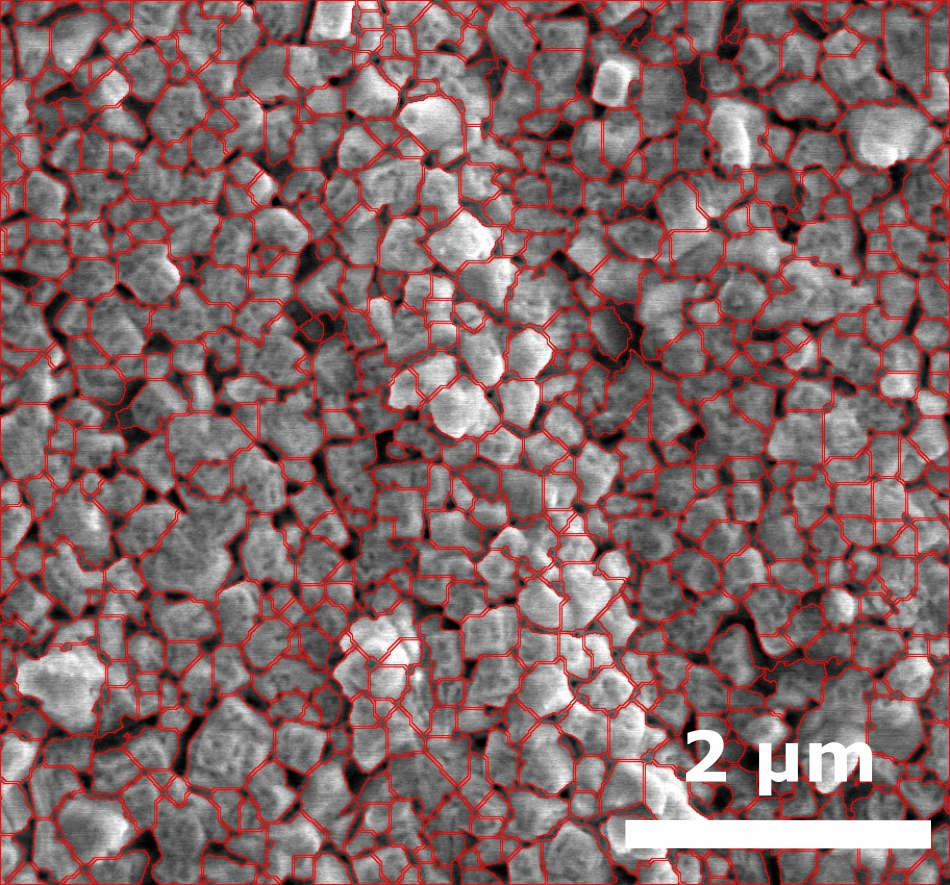}
\caption{An example of MIPAR processed image showing grain configurations on the electroplated sample.}\label{fig:Gfig}
\end{figure}

\begin{figure}[!h]
\includegraphics[width=0.37\textwidth]{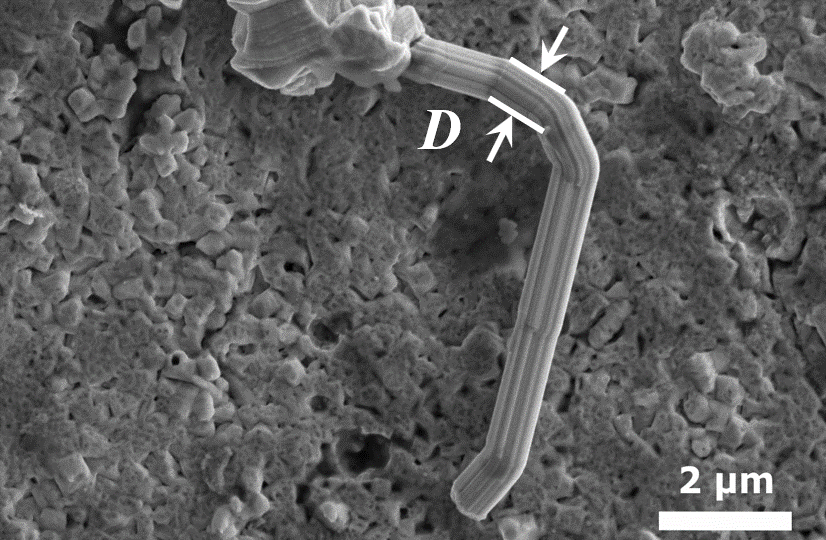}
\caption{An example of a MW diameter measurement \label{fig:WDmeasur}}
\end{figure}

\begin{figure}[!h]
\includegraphics[width=0.37\textwidth]{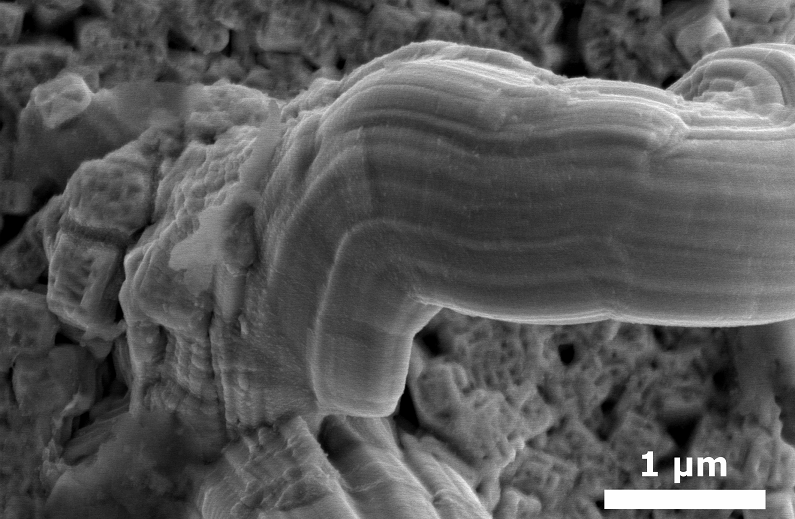}
\caption{An example of the composite MW and grain morphology of an electroplated sample showing how its development involves multiple grains much smaller than the MW diameter. }\label{fig:MWgrain}
\end{figure}

Our imaging results are illustrated in Figs. \ref{fig:Gfig}, \ref{fig:WDmeasur}, and \ref{fig:MWgrain}. The film surface was captured by using a scanning electron microscope (SEM), Hitachi S-4800 in a mixed secondary electron detector mode with an acceleration voltage of 5 kV, a magnification of 2 K, a working distance of 16.6 mm, and an e-beam current of 10 $\mu$A.


The Materials Image Processing and Automated Reconstruction (MIPAR)\cite{mipar1,mipar2} software package was used to collect the grain size statistics. Each MIPAR processed image required a recipe consisting of Wiener filtering \cite{mipar1}, adaptive thresholding, separating the grains, filling all holes, rejecting features, and calibrating the scale, as explained e. g. in Ref. \onlinecite{campbell2018}.

\begin{figure*}
\includegraphics[width=0.35\textwidth,]{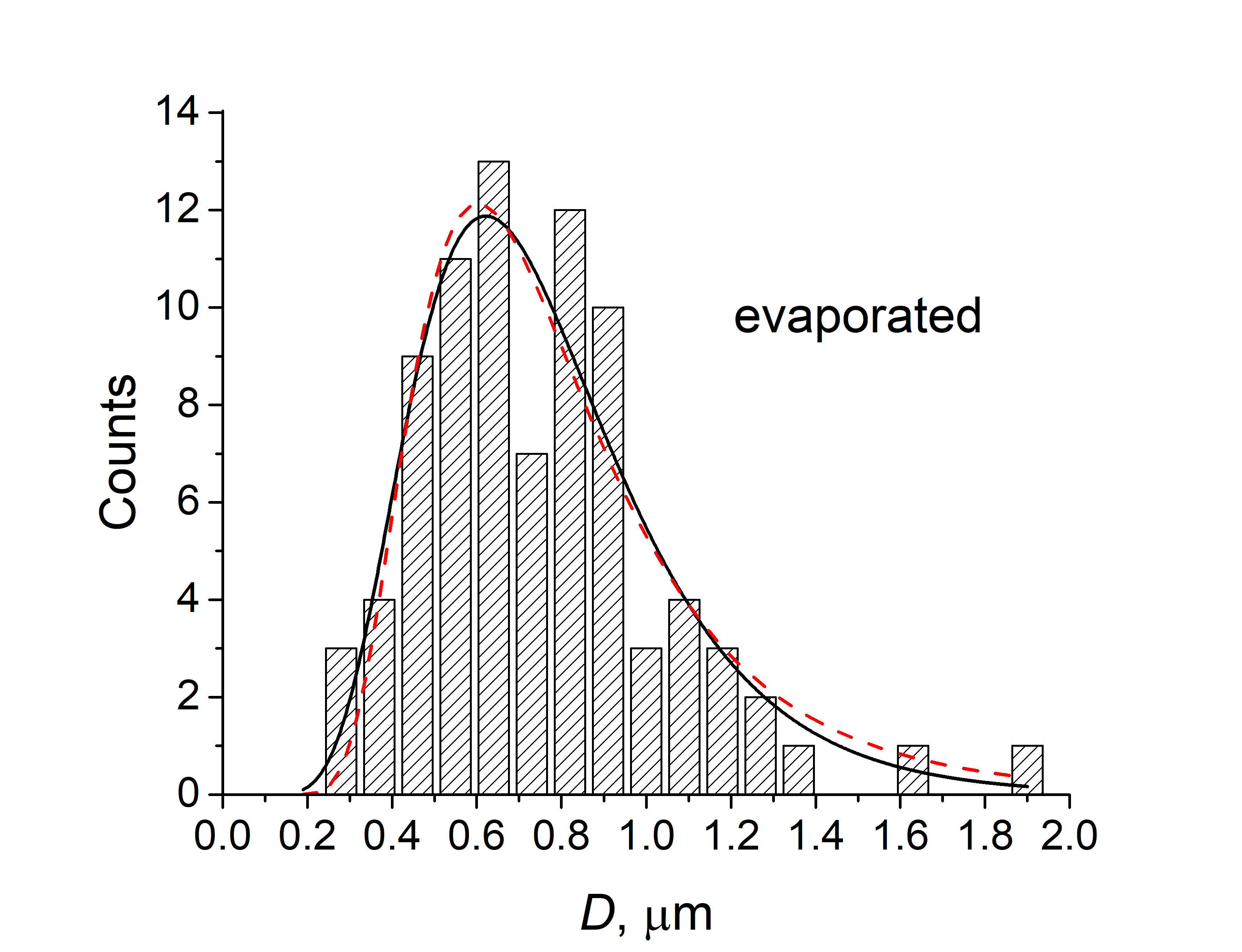}\includegraphics[width=0.35\textwidth,]{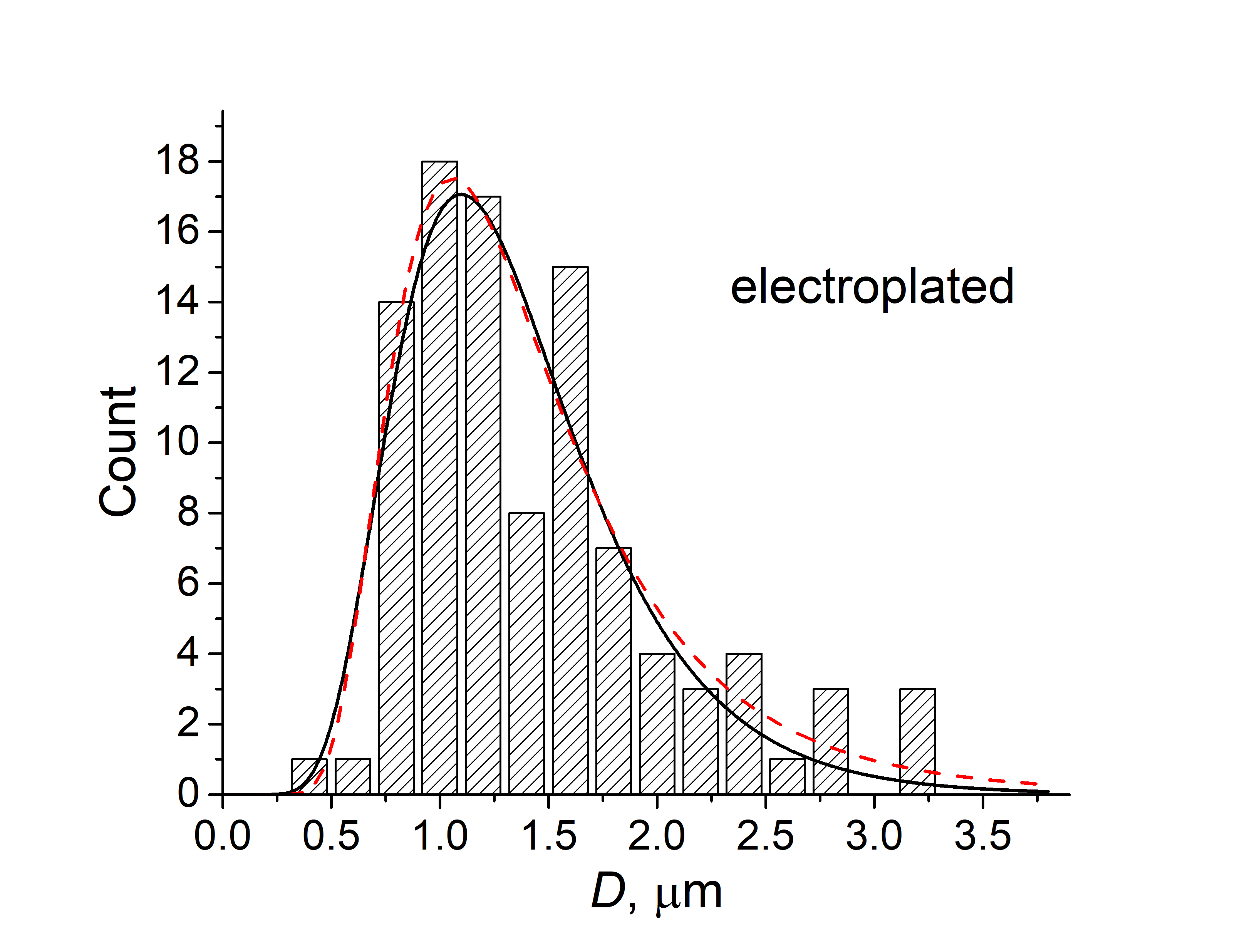}
\includegraphics[width=0.35\textwidth,]{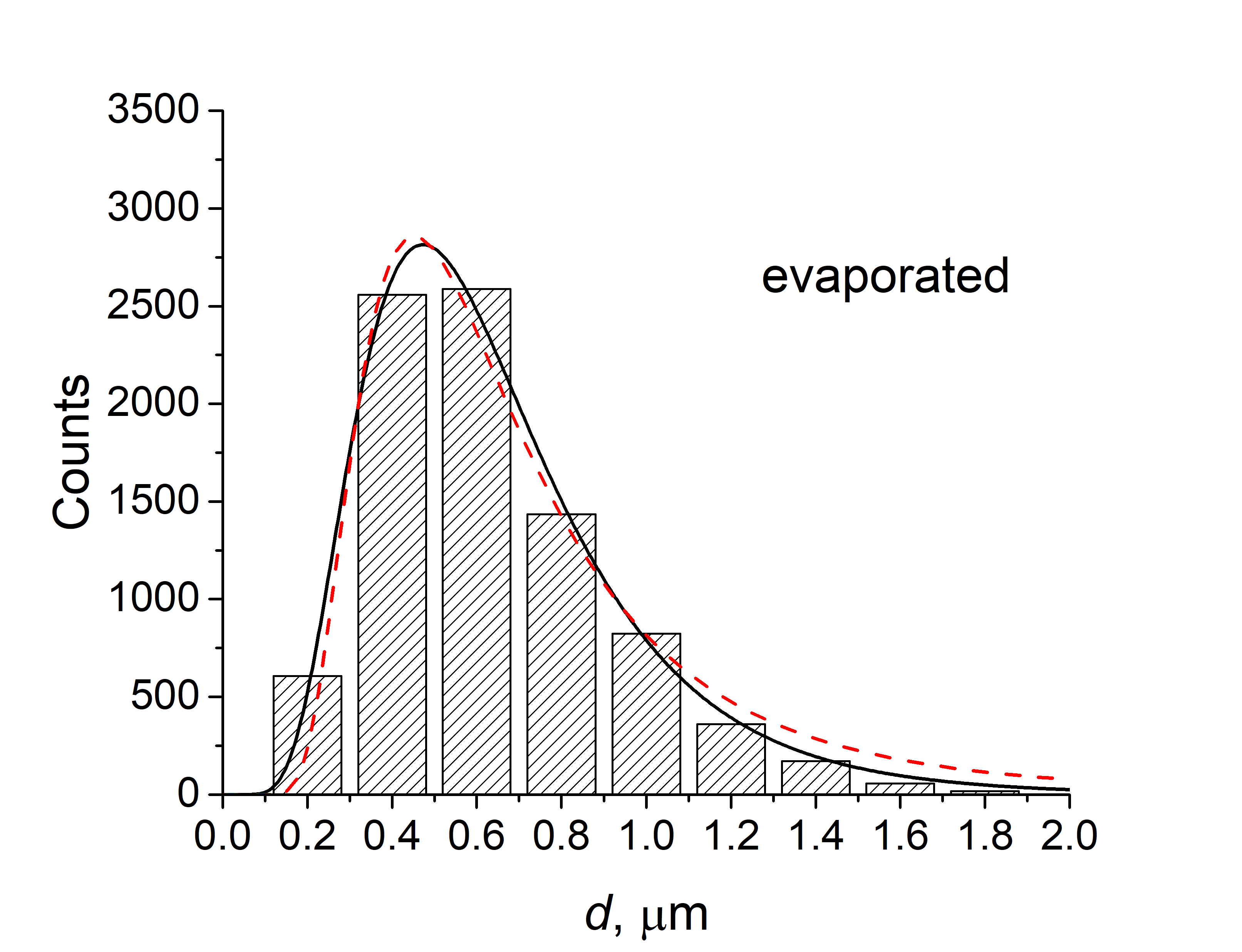}\includegraphics[width=0.35\textwidth,]{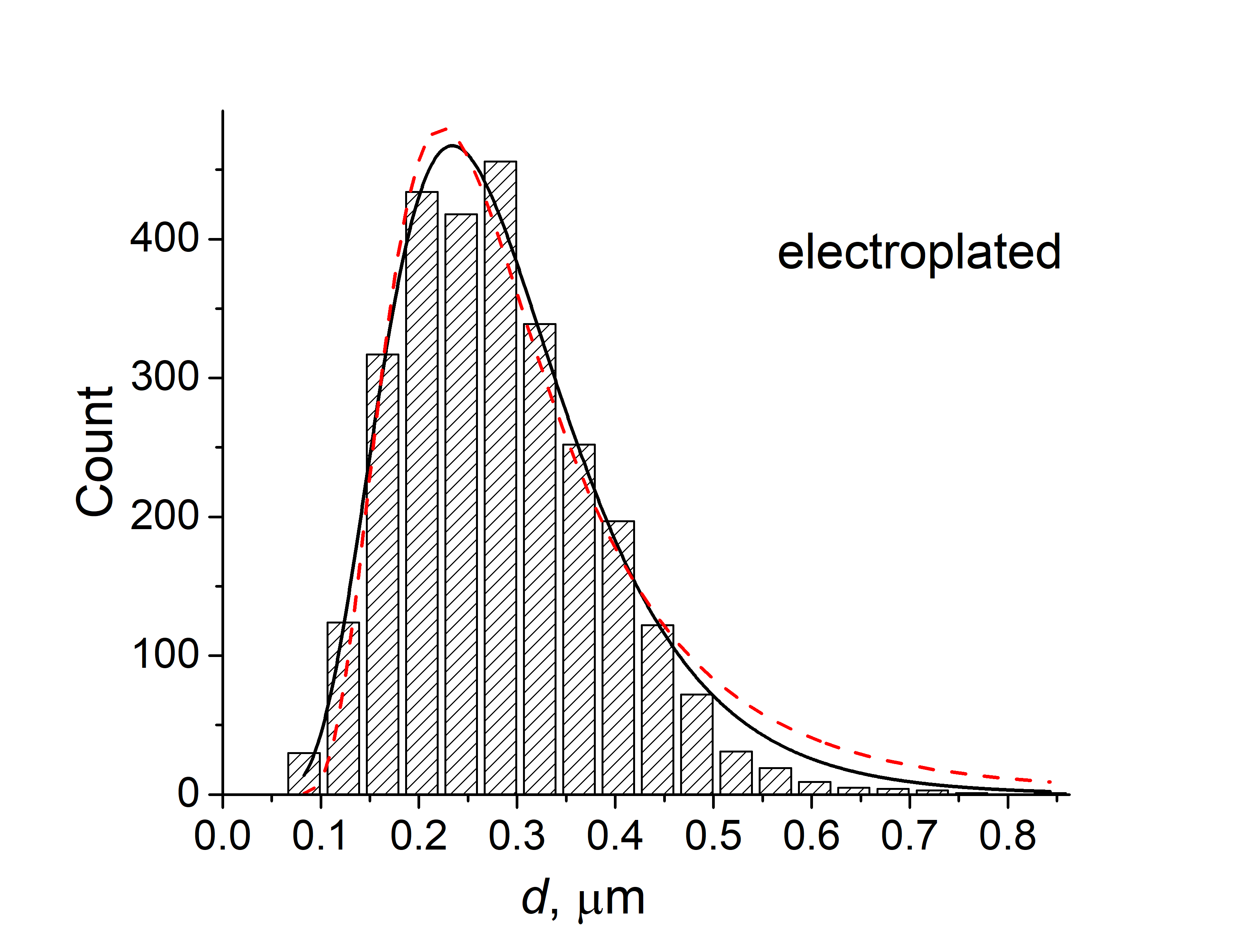}
\caption{The histograms and log-normal fits for the statistical distributions of whisker diameters ($D$; top row) and effective grain diameters ($d$; bottom row) for the evaporated and electroplated Sn films. Solid lines represent the log-normal fits by Eq. (\ref{eq:LNS}). Dashed lines represent the modified log-normal fits by Eq. (\ref{eq:LNSM}).}\label{fig:stats}
\end{figure*}

\begin{figure*}[t]
\includegraphics[width=0.35\textwidth,]{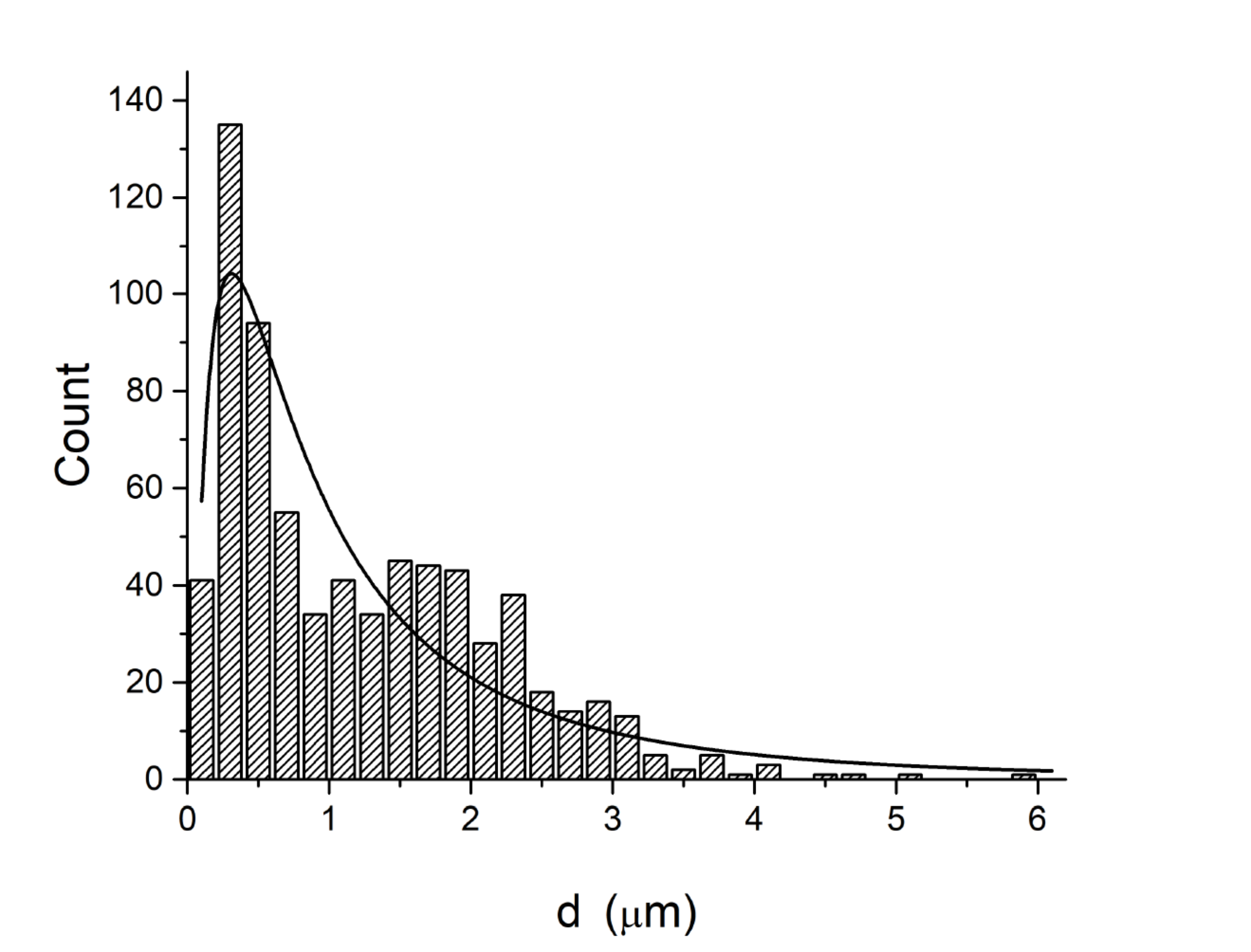}\includegraphics[width=0.35\textwidth,]{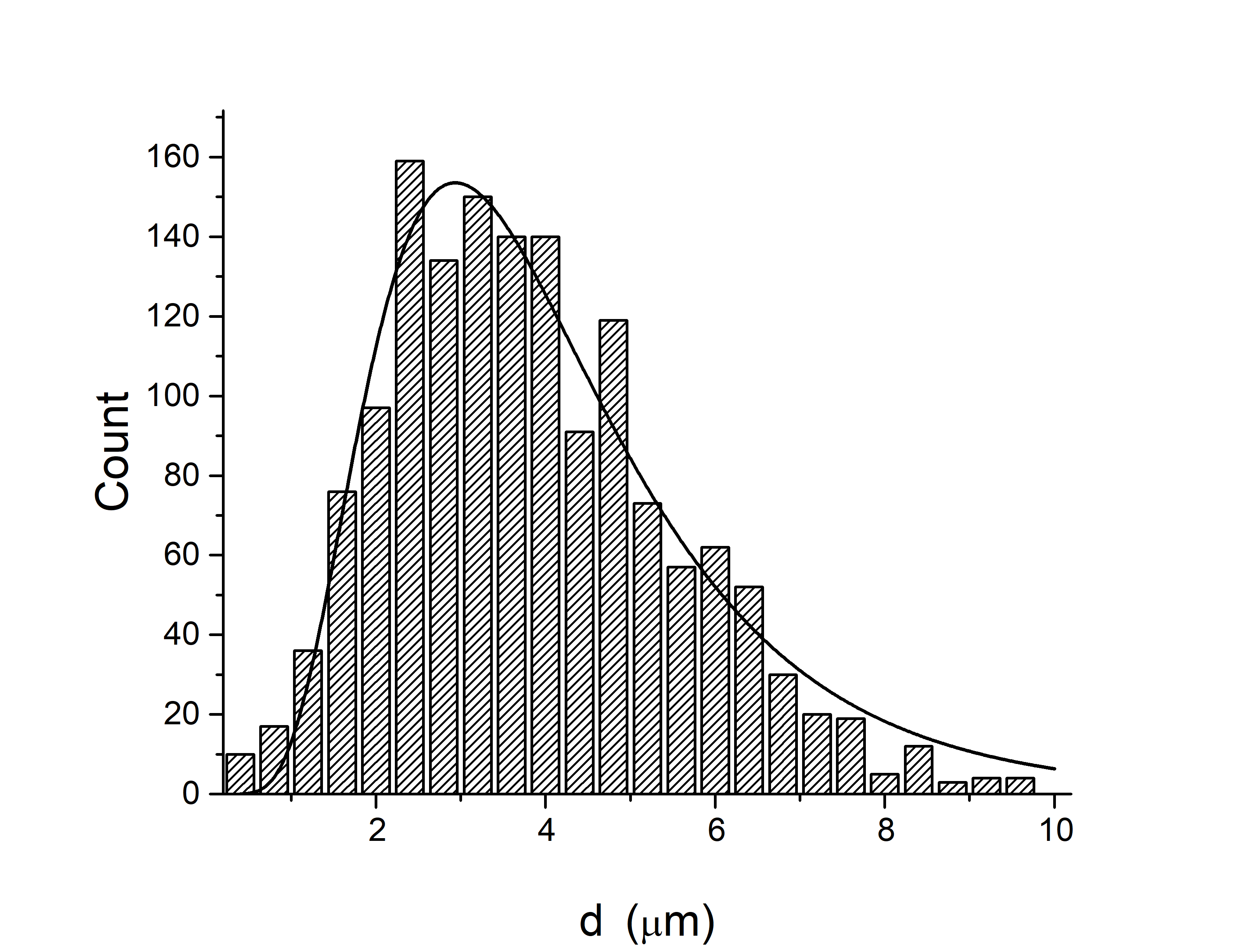}
\caption{The histograms and log-normal fits for the statistical distributions of effective grain diameters  for the electroplated Sn films; extracted from the data in Ref. \cite{jagtap2017}. Solid lines represent log-normal fits. }\label{fig:msu}
\end{figure*}

More specifically, the Wiener filter with window size of 5 pixels was used to remove most of the additive noise. The adaptive threshold constraints were window size with 32 pixels and 72  percentage. The watershed algorithm was used to separate the connected grains. The fill-all-holes feature was applied and the rejecting features selected to remove the areas less than or equal to 20 pixel. An example of the processed picture showing grains is presented in Fig. \ref{fig:Gfig}. We then collected the statistics of grain equivalent diameters defined as those of the same area circles.

100 images of area 0.0028 mm$^{2}$ were taken for each of the evaporated and electroplated samples to collect the statistics of MW diameters measured with ImageJ software. For each MW, the diameter was defined as the shortest distance between two visually parallel edges of its 2D image projection, as illustrated in Fig. \ref{fig:WDmeasur}.

In all cases, we ignored MW features of irregular significantly non-cylindrical shape such as close to the base of the MW in Fig. \ref{fig:WDmeasur}. Additionally, we collected the statistics for the diameters measured close to MW tips and midpoints: no significant differences were observed (although the diameter variations of several percent along MW lengths were typical). The uncertainty related to the MW cross sections deviating from the circular shape remains in our adopted diameter measurements taken from 2D images. That uncertainty is effectively tantamount to the approximation of  MW equivalent diameter defined as the same area circular cross section and similar to that adopted for grains.

\begin{table*}[!t]
\centering
\caption{Best fit parameters of log-normal distributions}
\begin{tabular}{|l|c|c|c|c|c|c|c|c|c|c|c|c|}
   \hline
Object &GEP$^1$ &WEP$^2$ & GEV$^3$ & WEV$^4$ & GEP1$^5$ &GEP2$^5$ & WEP1$^6$ &WEP2$^6$ & WEP3$^6$ & WEP4$^7$ & WEP5$^7$ & WEP6$^8$
\\ \hline
  $\mu$, $\mu$m & 0.27& 1.27 & 0.59 &0.71 &0.38 & 1.27 & 1.48 & 1.5& 1.46 &1.24 & 2.5 &1.17 \\  \hline
  $\Delta$ & 0.39 & 0.38 &0.47 & 0.38 & 0.36 & 0.38& 0.4 &0.43 & 0.37 & 0.46& 0.76 & 0.67\\ \hline
 \end{tabular}
\begin{tablenotes} \item [1] $^1$GEP stands for grains in our electroplated films. $^2$WEP stands for whiskers in our electroplated films. $^3$ GEV stands for grains in our evaporated films. $^4$WEV stands for whiskers in our evaporated films. $^5$The parameters are extracted from the fits of Fig. \ref{fig:msu}. $^6$The parameters WEP1,WEP2, WEP3 are from Ref. \onlinecite{panashchenko2009}. $^7$The parameters WEP4 and WEP5 are obtained by log-normal fitting of the data from Ref. \onlinecite{meschter2015}. $^8$The parameters WEP6 are from Ref. \onlinecite{panashchenko2012}. \end{tablenotes}
\label{tab:param}\end{table*}

To verify the representativeness of our data we have additionally extracted the grain diameter statistics from published work. \cite{jagtap2017} We applied the above described MIPAR procedure to the images in Figs. 3 and 10 of Ref. \onlinecite{jagtap2017} that correspond to the electroplated samples deposited under different conditions; the results are shown in Fig. \ref{fig:msu}. Our electroplated sample grain diameter distribution in Fig. \ref{fig:stats} is rather close to the one of Fig. \ref{fig:msu}.

An important additional feature pertaining  to both the grains and the MWs, and illustrated in Fig. \ref{fig:MWgrain}, is that multiple relatively small grains can be observed as part of the base of MW structure.

\section{Discussion}\label{sec:disc}
\subsection{Log-normal statistics}\label{sec:lns}

As is seen from Figs. \ref{fig:stats} and \ref{fig:msu}, all the statistics for MW and grain diameters can be fairly well fit with the log-normal distribution, \cite{lognorm}
\begin{equation}\label{eq:LNS}
\rho _X(X)=\frac{1}{\sqrt{2\pi\Delta} }\frac{1}{X}\exp\left[-\frac{(\ln X-\mu)^2}{2\Delta }\right] \end{equation}
where $X=d$ or $X=D$ represents respectively the grain or whisker diameter, $\mu$ and $\Delta$ are the mean and variance.

The best fit values of the latter parameters are given in Table \ref{tab:param}. It follows that for the electroplated samples, MW diameters, on average, significantly exceed that of grains, consistent with the morphology in Fig. \ref{fig:MWgrain}. That inequality is particularly  strong for our own data showing that if MW originate from individual grains {\it all obeying the same log-normal statistics}, then their concentration at $D\sim 3$ $\mu$m would be more than an order of magnitude lower than observed. We conclude that the hypothesis of individual MWs growing from single grain each would imply that MW underlying grains do not obey the observed log-normal statistics. On the other hand, assuming MW - single grain correspondence, the underlying grain diameter statistics must be log-normal following that of the observed whiskers. With the latter contradiction in mind, the MW - single grain one-to-one correspondence appears unlikely for the electroplated samples.

However, our data on the evaporated samples are consistent with the hypothesis of MW originating from a single grain each: MW and grain distribution parameters are close to each other. We thus arrive at the conclusion that attributing MWs to individual grains can be statistically justified in some, but not all cases. Certain morphologies, such as our and others \cite{kakeshita1982} electroplated samples point at MWs diameters on average covering several underlying grains.

Starting from the statistical `theory of breakage' \cite{kolmogorov1941,epstein1947} and developed turbulence \cite{kolmogorov1941,yaglom1966}, a consensus emerged that the log-normal statistics can originate from the multiplicative random processes, such as a rock disintegrating into $n_1$ random pieces, which, in their turn, disintegrate randomly into $n_2$ each, etc., or a turbulent flow generating random eddies that split into smaller ones in a self-similar manner. The total number $N$ of objects created in sequence of $M$ such processes is given by a product of random quantities, $N=\prod _{i=1}^Mn_i.$ Therefore, its logarithm becomes a sum of many random contributions, $\ln N=\sum _{i=1}^M\ln n_i$  obeying the central limit theorem; hence, the normal distribution for $\ln N$ of the type in Eq. (\ref{eq:LNS}).

Time reversal of the latter type of processes represents coalescence of random particles that can underly the observed log-normal distributions of grain diameters. \cite{granquist1976,pande1987} For example, small random islands created under deposition on a substrate, will coalesce into bigger ones, the latter randomly forming even bigger islands, etc. The final generation of mutually constrained grains is incapable of further coalescence with each grain consisting of much smaller co-grown crystallites.

Because the parameters of MWs log-normal distributions are comparable to that of grains, it is natural to assume that a similar coalescence process is responsible for the MW diameter distribution. That would imply that each MW is formed by co-grown thin metal filaments, the coalescence of which forms filaments of larger diameters, etc., until the process is terminated and a final MW is formed. 

Note that a scenario \cite{kakeshita1982} assuming the grain recrystallization with its diameter increasing in the process of MW growth, is not alternative to that put forward here. Our outlined understanding is that there is a factor (energy minimization) making neighboring grains to merge and form a single platform for a whisker. It remains uncertain in our scenario how such a merger takes place, except  that it decreases the electrostatic energy.

Remarkably, that same scenario was recently put forward in order to explain the observed evidence of multi-filament structure of MWs. \cite{borra2018} It is based on the electrostatic concept, according to which, MWs grow on local spots exhibiting significant enough surface charge density and its corresponding normal component of the electric field. The charging is related to various surface imperfections, such as a `wrong' grain orientations, contaminations, deformations, etc. In particular, several adjacent grains bearing significant charges can give rise to MWs of diameters exceeding that of individual grains. Therefore, the above scenario not only explains the similarity between the grain and MW statistics, but also elucidates the nature of MWs with relatively large diameters. It explains as well the imperfect MWs cross sections, and the presence of striations on their side surfaces. \cite{borra2018}

Earlier, it was observed indeed that the morphology of the whiskers was a result of whether they had nucleated on a single grain or on multiple grains. In the latter case, the whisker surface was fluted or striated. A whisker formed by nucleation from several grains that surrounded a region of porosity could result in a hollow whisker. \cite{lebret2003} A more recent study \cite{michael2012} stated the lack of simple relationship between the predominant crystallographic whisker growth directions and the film texture with whiskers generally growing from grains that did not correspond to the major textures in the film, but otherwise can have various orientations.

\subsection{Modified log-normal statistics}\label{sec:lnsm}
One uncomfortable feature of the log-normal distribution in Eq. (\ref{eq:LNS}) is that the physical meaning of the dispersion parameter $\Delta$ remains unknown. Revisiting the above outlined justification of log-normal statistics, one can notice that the dispersion of quantity $\ln N=\sum _i^M\ln n_i$ should be estimated as
\begin{equation}\label{eq:DM}\Delta _M=M\langle [\delta\ln( n_i)]^2\rangle \end{equation}
where $\langle [\delta\ln( n_i)]^2\rangle$ represents the sub-processes dispersion. Because $\Delta _M$ is the dispersion that is required by the central limit theorem, it should be used in place of $\Delta$ in Eq. (\ref{eq:LNS}). The proportionality of dispersion to the number of cycles $M$, overlooked in the previously proposed justifications, \cite{granquist1976,pande1987,vaz1988,kiss1999} is taken into account here.

The total number of coalescence cycles involved can be estimated from the increase in cross-sectional areas, $X_M^2=X_0^2k^M$. Here $X_0$ is the `elemental' (initial) size of the object in the beginning of coalescence, and $k$ is the time reversed multiplication factor, $n_i=n_{i-1}/k$. Combining these estimates yields,
\begin{equation}\label{eq:M}
M=\frac{2\ln(X/X_0)}{\ln k}\end{equation}
and
\begin{equation}\label{eq:alpha}
\Delta _M=2\alpha\ln(X/X_0)\quad {\rm with}\quad \alpha =\frac{\langle [\delta\ln(n_i)]^2\rangle }{\ln k}. \end{equation}
Using $\Delta _M$ with Eq. (\ref{eq:LNS}) and neglecting the logarithmic dependence in its pre-exponential factor, modifies the log-normal distribution to the form,
\begin{equation}\label{eq:LNSM}
\rho _X(X)\propto\frac{1}{X}\exp\left[-\frac{(\ln X-\mu)^2}{4\alpha\ln(X/X_0)}\right]. \end{equation}

When a particle (grain or whisker) diameter is not too far from the distribution maximum, the modified form in Eq. (\ref{eq:LNSM}) is fairly close to the original one in Eq. (\ref{eq:LNS}) because the logarithm in the denominator of the exponent is a slow function that can be approximated with a constant when $X\sim \exp(\mu )\gg X_0$. However, the distribution in Eq. (\ref{eq:LNSM}) falls faster towards small diameters and it decays slower towards large diameters.

In the limiting case of $X\gg \exp(\mu )$ and $X\gg X_0$, Eq. (\ref{eq:LNSM}) reduces to a power type dependence, $\rho\propto X^{-(1+1/4\alpha)}$. One practical consequence of the latter prediction is that the concentration of `dangerously' thick MWs (capable of punching through insulating layers) may not be exponentially small thus aggravating the reliability concerns.

Eq. (\ref{eq:LNSM}) clarifies the nature of log-normal dispersion and can be used to estimate it. We evaluate $\alpha$ in Eq. (\ref{eq:LNSM}) from the small value relation
\begin{equation}\label{eq:deltalog}\delta \ln(n_i)=\overline{\delta n_i}/\langle n_i\rangle\approx 1/\sqrt{\langle n_i\rangle}\end{equation}
where, $\overline{\delta n_i}\equiv \sqrt{\langle(\delta n_i)^2\rangle}$. In order to discriminate between two consecutive cycles, the scale factor $k$ determining the average change in particle number per cycle,
\begin{equation}\label{eq:key}\langle n_{i-1}-n_i\rangle =(k-1)\langle n_i\rangle\end{equation}
must be noticeably, by a certain numerical factor $\beta$, greater than its fluctuation; hence,
\begin{equation}\label{eq:beta}
(k-1)\langle n_i\rangle \approx \beta\overline{\delta n_i}.\end{equation}
Expressing from here the ratio $\overline{\delta n_i}/\langle n_i\rangle$, combining with Eq. (\ref{eq:deltalog}), and assuming small $k-1\ll 1$, yields
\begin{equation}\label{eq:alpha1}\alpha=\frac{(k-1)^2}{\beta ^2\ln k}=\frac{1}{\beta ^2}.\end{equation}

The numerical coefficient $\beta$ introduced in Eq. (\ref{eq:beta}) and independent of the number of cycles remains unknown. Based on its meaning as the minimum number of standard deviations necessary to discriminate between two overlapping distributions, one can expect $\beta \sim 2-3$.

The  data in Fig. \ref{fig:stats} can be fit with Eqs. (\ref{eq:LNS}) and Eqs. (\ref{eq:LNSM}) almost equally well, with Eq. (\ref{eq:LNS}) working slightly better for the case of grains while Eq. (\ref{eq:LNSM}) better fitting the distributions for whiskers. In our fitting procedure with Eq. (\ref{eq:LNSM}), we have used $X_0=50$ nm corresponding to the characteristic size of Sn film crystallite as obtained from the X-ray diffraction. \cite{borra2018} As a result, our best fit parameters $\alpha$ corresponding to Fig. \ref{fig:stats} fall in the interval of $\alpha\sim 0.1-0.15$, consistent with the estimate in Eq. (\ref{eq:alpha1}).

\section{Conclusions}\label{sec:conl}
Based on our collected statistics of tin grain and whisker diameters we conclude the following:\\
(i) Both statistics are well fit with the log-normal distributions. \\
(ii) The parameters of those distributions are not dramatically different and are consistent with other published work that are available. \\
(iii) MW diameters in electroplated samples are systematically larger than that of grains.\\
(iv) The distributions of MW and grain diameters in evaporated samples are close to each other.\\
(v) The observed log-normal statistics and the fact that MW diameters can exceed those of grains are consistent with the concept of multiple filament structure of MWs. \cite{borra2018}\\
(vi) A  modification of particle size log-normal distribution is developed clarifying the nature and size dependence of its dispersion.\\
(vii) On a practical side, our results predict the probability of growing thick whiskers presenting elevated hazard for electronic package reliability. \\

\section*{Acknowledgement}
We are grateful to D. Shvydka and R. Irving for useful discussions and encouragement and to D. Niraula for proofreading the manuscript. Also, we grateful to  S. Smith for useful discussions and for bringing to our attention Ref. \onlinecite{michael2012}.


\begin{thebibliography}{99}
\bibitem{NASA1}NASA Goddard Space Flight Center Tin Whisker Homepage, website \url{http://nepp.nasa.gov/whisker}.
\bibitem{barnes}J. R. Barnes, Bibliography for Tin Whiskers, Zinc Whiskers, Cadmium Whiskers, Indium Whiskers, and Other Conductive Metal and Semiconductor Whiskers; \url{http://www.dbicorporation.com/whiskbib.htm}
\bibitem{galyon2003}G. T. Galyon, Annotated Tin Whisker Bibliography And Anthology, IEEE Transactions on electronics Packaging Manufacturing, {\bf 28}, 94 (2005); \url{http://thor.inemi.org/webdownload/newsroom/TW_},\url{biblio-July03.pdf}

\bibitem{brusse2002}J. Brusse, G. Ewell, and J. Siplon, Tin Whiskers: Attributes and Mitigation, Capacitor and Resistor
Technology Symposium (CARTS), March 25-29,  pp. 68-80, (2000).
\bibitem{panashchenko2009}L. Panashchenko, Evaluation of environmental Tests for tin whisker assessment, MS Thesis, University of Maryland (2009). http://hdl.handle.net/1903/10021
\bibitem{panashchenko2012}L. Panashchenko, The Art of Metal Whisker Appreciation, IPC Tin Whisker Symposium, Dallas, TX, 2012,
\burl{https://nepp.nasa.gov/whisker/reference/tech_papers/2012-Panashchenko-IPC-Art-of-Metal
\bibitem{borra2018}V. Borra, D. G. Georgiev, V. G. Karpov, and D. Shvydka, Microscopic structure of metal whiskers, Phys. Rev. Applied, {\bf 9}, 054029 (2018).

\bibitem{karpov2014} V. G. Karpov, Electrostatic theory of metal whsikers, Phys. Rev. Applied, {\bf 1}, 044001  (2014).
\bibitem{karpov2015}V.G.Karpov, Electrostatic Mechanism of Nucleation
and Growth of Metal Whiskers, SMT Magazine, February 2015, p. 28. \url{http://iconnect007.uberflip.com/i/455818/44}
\bibitem{fang2006}T. Fang, M. Osterman, M. Pecht, Statistical Analysis of Tin Whisker Growth, Microelectronics Reliability {\bf 46}, 846 (2006).


\bibitem{susan2013}D. Susan, J. Michael, R. P. Grant, B. McKenzie \& W. G. Yelton, Morphology and Growth Kinetics of Straight and Kinked
Tin Whiskers, Metall and Mat Trans A, {\bf 44}, 1485 (2013).
\bibitem{meschter2015}S. Meschter and P. Snugovsky, “Tin whisker testing and modeling,” SERDP Project WP-1753, Final Report No. PM-LF-2015-11, 2015,
\url{https://www.serdp-estcp.org/Program-Areas/Weapons-Systems-and-Platforms/Lead-Free-Electronics/WP-1753}.

\bibitem{niraula2015}D. Niraula and V. G. Karpov, The probabilistic distribution of metal whisker lengths, J. Appl. Phys. {\bf 118}, 205301 (2015).
\bibitem{subedi2017}B. Subedi, D. Niraula, and V. G. Karpov, The stochastic growth of metal whiskers, Appl. Phys. Lett. {\bf 110}, 251604 (2017); doi: 10.1063/1.4989852
\bibitem{davy2014}G. Davy, private communication, 10/2014; quated in \cite{karpov2015}.

\bibitem{zhang2004}Y. Zhang, Tin Whisker Discovery and Research, in {\it Soldering in Electronics}, Edited by K. Suganuma, Marcel Dekker, Inc. p. 121 (2004)
\bibitem{tu2005}K.N. Tu, J.O. Suh, and Albert T. Wu, Tin Whisker Growth on Lead-Free Solder Finishes, in {\it Lead-Free Solder Interconnect Reliability}, Edited by D. Shangguan, ASM International, p. 851 (2005).
\bibitem{bunian2013}D. Bunyan, M. A. Ashworth, G. D. Wilcox, R. L. Higginson, R. J. Heath, C. Liu, Tin whisker growth from electroplated finishes – a review, Transactions of the institute of metal finishing, {\bf 91}, 249-259 (2013).

\bibitem{jagtap2017}P. Jagtap, A. Chakraborty, P. Eisenlohr, P. Kumar, Identification of whisker grain in Sn coatings by analyzing
crystallographic micro-texture using electron back-scatter diffraction, Acta Materialia {\bf 134}, 346 (2017).
\bibitem{pei2012}F. Pei, N. Jadhav, and E. Chason, Correlating whisker growth and grain structure on Sn-Cu samples by real-time
scanning electron microscopy and backscattering diffraction characterization, Appl. Phys. Lett. {\bf 100}, 221902 (2012); doi: 10.1063/1.4721661.
\bibitem{kakeshita1982}T. Kakeshita, K. Shimizu, R. Kawanaka, T. Hasegawa, Grain size effect of electro-plated tin coatings on whisker growth, J. Materials Science, {\bf 17}, 2560 (1982).
\bibitem{granquist1976}C. G. Granqvist and R. A. Buhrman, Ultrafine metal particles, J. Appl. Phys., {\bf 47}, 2200 (1976); doi: 10.1063/1.322870
\bibitem{pande1987}C. S. Pande, On a stochastic theory of grain growth, Acta memll. {\bf 35}, 2671, (1987).
\bibitem{vaz1988}M.F\`{a}tima Vaz and M.A.Fortes, Grain size distribution: the lognormal and the gamma distribution functions, Scrlpta Metallurgica, {\bf 22}, 35 (1988).

\bibitem{kiss1999}L. B. Kiss, J. S.Huderlund, G. A. Niklasson and C. G. Granqvist, New approach to the origin of lognormal size distributions of nanoparticles, Nanotechnology, {\bf 10}, 25 (1999).
\bibitem{vasko2015a}A. C. Vasko, C. R. Grice, A. D. Kostic, and V. G. Karpov, Evidence of electric-field-accelerated growth of tin whiskers, MRS Communications, Materials Research Society, (2015), doi:10.1557/mrc.2015.64 .
\bibitem{killefer2017}M. Killefer, V. Borra, A. Al-Bayati, D. G. Georgiev, V. G. Karpov, E. I. Parsai, and D. Shvydka, Whisker growth on Sn thin film accelerated under gamma-ray induced
electric field, J. Phys. D: Appl. Phys. {\bf 50}, 405302 (2017).
\bibitem{borra2016}V. Borra, D. G. Georgiev, and C.R. Grice, Fabrication of optically smooth Sn thin films, Thin Solid Films, {\bf 616}, 311-315 (2016).
\bibitem{mipar1}MIPAR software developer homepage; website \url{http://www.mipar.us/}
Access Volume 21, Issue S3 ( 2015) August 2015 , pp. 455-456
\bibitem{mipar2}J.M. Sosa , D.E. Huber, B.A. Welk, and H.L. Fraser, MIPAR: 2D and 3D Microstructural Characterization Software Designed for Materials Scientists, by Materials Scientists, Proceedings of Microscopy \& Microanalysis,{\bf 21}, 455 (2015); doi: 10.1017/S1431927615003074
\bibitem{campbell2018}A. Campbell, P. Murray, E. Yakushina, S. Marshall, W. Ion, New methods for automatic quantiﬁcation of microstructural featuresusing digital image processing, Materials \& Design
{\bf 141}, 395 (2018)
\bibitem{lognorm}{\it Lognormal distributions : theory and applications}(Series: Statistics, textbooks and monographs v.8), Edwin L. Crow and Kunio Shimizu, Eds. New York: M. Dekker, 1988.

\bibitem{kolmogorov1941}A.N.Kolmogorov, On the log-normal distribution of particles sizes during break-up process. Dokl. Akad. Nauk. SSSR {\bf 31}, 99 (1941).
\bibitem{epstein1947}B. Epstein, The mathematical description of certain breakage mechanisms leading to logarithmico normal distribution, J. Franklin Inst. {\bf 244},471 (1947).
\bibitem{yaglom1966}A. M. Yaglom, Fluctuations in energy dissipation as influencing the shape of turbulence characteristics in an inertial interval, Dokl. Akad. Nauk SSSR, {\bf 166}, 49 (1966) [Sov. Phys.-Dokl. {\bf 11} ,26 (1966)].
\bibitem{lebret2003}J.B. LeBret and M.G. Norton, Electron microscopy study of tin whisker growth, J. Mater. Res., {\bf 18}, 585 (2003).
\bibitem{michael2012}J. R. Michael, B. B. McKenzie and D. F. Susan, Application of Electron backscatter diffraction to teh crystallographic characterization of tin whiskers, in: {\it Understanding and Predicting Metallic Whisker Growth and its Effects on Reliability: LDRD Final Report}, SANDIA REPORT
SAND2012-0519, Unlimited Release, Printed January 2012, p. 42, available at \url{http://prod.sandia.gov/techlib/access-control.cgi/2012/120519.pdf}

\end{thebibliography}
\end{document}